\documentclass[runningheads]{llncs}
\usepackage{graphicx}
\usepackage{bbding}
\usepackage[colorlinks,linkcolor=blue, anchorcolor=blue, citecolor=blue, urlcolor=blue]{hyperref} 
\usepackage{enumerate} 
\usepackage{multirow} 
\usepackage{relsize} 
\usepackage{amsmath} 
\usepackage{amssymb} 
\usepackage{booktabs} 
\usepackage{svg} 

\begin{document}
\title{One Network to Solve Them All: A Sequential Multi-Task Joint Learning Network Framework for MR Imaging Pipeline}
%
%
\author{Zhiwen Wang\inst{1} \and
Wenjun Xia\inst{1} \and Zexin Lu\inst{1} \and Yongqiang Huang\inst{1} \and Yan Liu\inst{2} \and Hu Chen\inst{1} \and Jiliu Zhou\inst{3}
\and Yi Zhang\inst{1(}\Envelope\inst{)}}
\authorrunning{Z. Wang et al.}
\titlerunning{SemuNet:A Sequential Multi-task Joint Learning Network Framework }
%
\institute{College of Computer Science, Sichuan University, Chengdu 610065, China
\email{yzhang@scu.edu.cn} \and
College of Electrical Engineering, Sichuan University, Chengdu 610065, China \and
Chengdu University of Information Technology, Chengdu 610225, China
}
%
\maketitle              
\begin{abstract}
Magnetic resonance imaging (MRI) acquisition, reconstruction, and segmentation are usually processed independently in the conventional practice of MRI workflow. It is easy to notice that there are significant relevances among these tasks and this procedure artificially cuts off these potential connections, which may lead to losing clinically important information for the final diagnosis. To involve these potential relations for further performance improvement, a sequential multi-task joint learning network model is proposed to train a combined end-to-end pipeline in a differentiable way, aiming at exploring the mutual influence among those tasks simultaneously. Our design consists of three cascaded modules: 1) deep sampling pattern learning module optimizes the $k$-space sampling pattern with predetermined sampling rate; 2) deep reconstruction module is dedicated to reconstructing MR images from the undersampled data using the learned sampling pattern; 3) deep segmentation module encodes MR images reconstructed from the previous module to segment the interested tissues. The proposed model retrieves the latently interactive and cyclic relations among those tasks, from which each task will be mutually beneficial. The proposed framework is verified on MRB dataset, which achieves superior performance on other SOTA methods in terms of both reconstruction and segmentation. The code is available online: \url{https://github.com/Deep-Imaging-Group/SemuNet}

\end{abstract}
\section{Introduction}
Magnetic resonance imaging (MRI) is a non-invasive diagnostic imaging technique that enables studying low-contrast soft tissue structures without harmful radiation risk. However, its long acquisition time results in increasing costs, patient uncomfortableness, and motion artifacts. To conquer these obstacles, fast MRI acquisition is of great emergency. Nevertheless, simply reducing the sampling rate will degrade the imaging quality and jeopardize the sequential diagnosis. In the past decades, numerous efforts have made to recover high-quality MR images from undersampled $k$-space data, e.g., compressed sensing (CS) and later deep learning based methods. In spite of fruitful results obtained, two defects can be sensed: 1) current undersampling patterns are empirically handtailored, e.g., radial, Cartesian, or Gaussian, which ignore the fact that different images may be suitable for different undersampling patterns; 2) suboptimal sampling pattern will lead to suboptimal reconstruction and finally impact the sequential analysis task. In summary, isolatedly handling the main steps in the whole imaging pipeline reveals the potential fact that both radiologists and computer aided intervention systems may be working with suboptimal reconstructed images. 

Recently, in the field of signal processing, task driven methods, which directly train an end-to-end network and neglects the explicit intermediate result, have drawn increasingly attention. For examples, Bojarski et al. trained a self-driving network which learns commands directly from cameras without recogniting any landmarkers~\cite{ref_article1}. Liu et al. proposed to integrate the denoising network with a segmentation network to improve the denoising performance for segmentation~\cite{ref_article2}. Encouraged by these promising results, similar ideas were introduced into the field of medical imaging. Wu et al. and Lee et al., respectively proposed to detect the pulmonary nodules and intracranial hemorrhage directly from the measured data without the step of image reconstruction~\cite{ref_article3,ref_article4}. In~\cite{ref_article5,ref_article6}, the authors coupled MRI reconstruction with segmentation to improve the performance of both tasks. On the other hand, some recent studies attempted to optimize the $k$-space sampling patterns with a data-driven manner~\cite{ref_article7,ref_article8,ref_article9} and the optimized undersampling patterns show significant improvements, compared to empirical ones. Unfortunately, the scheme of these trajectories only learned from the reconstruction stage ignore the tissue of interest. Meanwhile, these methods mentioned above either combine the sampling and reconstruction, or joint the tasks of reconstruction and segmentation. None of the existing works consider the whole pipeline of medical image analysis, which means that useful information for final segmentation may be lost in each step.

To fully explore the mutual influence among sequential tasks and further improve the performance of each task simultaneously, in this study, we propose a sequential multi-task joint learning network framework (SemuNet), which jointly optimizes the sampling, reconstruction and segmentation in an end-to-end manner. The proposed framework can be divided into three modules: the sampling pattern learning network (SampNet), the reconstruction network (ReconNet), and the segmentation network (SegNet). Specifically, the well-known U-Net~\cite{ref_article10} is adopted as the backbone of our proposed ReconNet and SegNet for simplicity,  and a probabilistic sampling network is proposed to learn the sampling pattern. 

The remainder of this paper is organized as follows. The details of the proposed model, including each module, are elaborated in Section 2. The experimental results are presented and discussed in Section 3 and the final section concludes this paper.

\section{Method}
In this section, the main modules of the proposed framework SemuNet, including SampNet, ReconNet and SegNet, are first described sequentially in detail. Then other issues of the SemuNet, especially about the training strategy and loss function, are presented.
\subsection{SampNet: the sampling pattern learning network}
For the problem of CS-MR imaging, the task is to reconstruct an MR image from undersampled measurements
in $k$-space, which approximates a fully-sampled MRI image $\textbf{x} \in \mathbb{C}^{\sqrt{N} \times \sqrt{N}}$. Let ${S}_{\textbf{T}_{c}}(\cdot)$ denotes the SampNet parameterized by $\textbf{T}_c$, which outputs a $\sqrt{N} \times \sqrt{N}$ continuous value matrix (i.e., sampling pattern) as a partial observation in $k$-space. The undersampling process can be written as $S_{\textbf{T}_c}(\textbf{x})\odot\textbf{F}\textbf{x}$ , where $\odot$ is Hadamard product, $\textbf{F}$ is the Fourier transform matrix. The goal of SampNet is to optimize the sampling pattern for specific datasets in the $k$-space. To learn a probabilistic observation matrix $\textbf{T}_c$ in the $k$-space, we adopt the similar architecture to the~\cite{ref_article8,ref_article10,ref_article12} for our SampNet. The architecture of SampNet is shown in Fig.~\ref{fig1}a. The details of SampNet is given in the supplementary material.

Since we do not have the labels for sampling pattern learning, we propose to merge the SampNet into the ReconNet and SegNet. When the cascaded network converges, the top-$n$ largest values in $\textbf{T}_c$ are replaced by Boolean values to produce the final sampling pattern $\textbf{T}$, and $n$ is chosen according to the predetermined sampling rate $\alpha$ , where $n=\alpha\cdot N$. Accordingly, the Booleanizing operation can be written defined as: 
\begin{equation}
(\textbf{T})_{ij} 
= \begin{cases}
1, & \text {if } (S_{\textbf{T}_c}(\textbf{x}))_{ij} \ \text{is in top-}{n}, \\
0, &\text{otherwise}
\end{cases}
\end{equation}
As a result, the pattern is optimized by the knowledge of both high-quality reconstructed images and accurate segmentation labels.

\subsection{ReconNet: the reconstruction network}
Recently, extensive network models were proposed for MRI reconstruction~\cite{ref_article13}, and in this work, we simply utilize the spatial-domain based reconstruction network. Letting $R_\theta(\cdot)$ denote the ReconNet with parameter set $\theta$, $\textbf{F}^{-1}$ is the inverse Fourier transform matrix, the reconstructed image $\tilde{\textbf{x}}$ can be obtained as:
\begin{equation}\label{eq5}
\tilde{\textbf{x}} =
R_{\theta}(\textbf{F}^{-1}S_{\textbf{T}_c}(\textbf{x})\odot\textbf{F}\textbf{x})
\end{equation}

Then the training procedure can be formulated as the following optimization problem:
\begin{equation}\label{eq6}
\{\theta^*, \textbf{T}_c^*\} = 
\underset{\theta, \textbf{T}_c}{\arg \min} {\mathbb{E}}_{\textbf{x}}
[g(R_{\theta}(\textbf{F}^{-1}S_{\textbf{T}_c}(\textbf{x})\odot\textbf{F}\textbf{x}), \textbf{x})]
\end{equation}
where $g(\cdot)$ is a reconstruction metric function to measure the similarity between the reconstructed image and the label, and $\mathbb{E}_\textbf{x}$ is the expectation over $\textbf{x}$.
The architecture of ReconNet adopts the well-known U-Net~\cite{ref_article10} as the backbone as shown in Fig.~\ref{fig1}b, which has demonstrated competitive performance in artifact reduction for MRI~\cite{ref_article8,ref_article14}.
\subsection{SegNet: the segmentation network}
Recently, lots of networks were proposed for automatic tissue segmentation~\cite{ref_article15}. Since U-Net like architecture has demonstrated excellent performance for medical image segmentation, in this part, we also choose the same network structure in Fig.~\ref{fig1}b as our SegNet for simplicity. Then we can formulate the joint learning for simultaneously optimizing sampling, reconstruction and segmentation as follows
\begin{equation}\begin{split}
\{\epsilon^*,\theta^*, \textbf{T}_c^* \} = 
\underset{\epsilon,\theta, \textbf{T}_c}{\arg \min} {\mathbb{E}}_{\textbf{x}}
[&g(R_{\theta}(\textbf{F}^{-1}S_{\textbf{T}_c}(\textbf{x})\odot\textbf{F}\textbf{x}), \textbf{x}) \\
+&h(H_{\epsilon}(R_{\theta}(\textbf{F}^{-1}S_{\textbf{T}_c}(\textbf{x})\odot\textbf{F}\textbf{x})),\textbf{s})]
\end{split}\end{equation}
where $h(\cdot)$ is the segmentation metric function to measure the segmentation accuracy of the result compared to the segmentation labels and $H_{\epsilon}(\cdot)$ is the segmentation network with parameter set $\epsilon$.

The SegNet plays two roles. First, it is treated as a clinical analysis instructor to train ReconNet, such that the reconstruction network can better adapt to tissue segmentation work. Second, it serves as a radiologist, which can provide SampNet with sufficient clinical knowledge.

\subsection{SemuNet: the sequential multi-task joint learning network framework}
By cascading the previously mentioned SampNet, ReconNet and SegNet as the basic modules, we propose a deep joint learning framework for the whole MRI pipeline, which can:
1). learn an optimized sampling pattern simultaneously guided by both low- and high-level tasks, i.e. reconstruction and segmentation;
2). reconstruct high-quality MR images with the optimized sampling pattern for the downstream segmentation task;
3). and segment the target tissues more accurate based on the task-driven reconstruction.

Since these modules are cascaded and trained in an end-to-end manner, the features extracted from different tasks are mutually influenced in an interactive way and benefit from each other. 
\begin{figure}
\begin{center}
\includegraphics[width=0.8\textwidth]{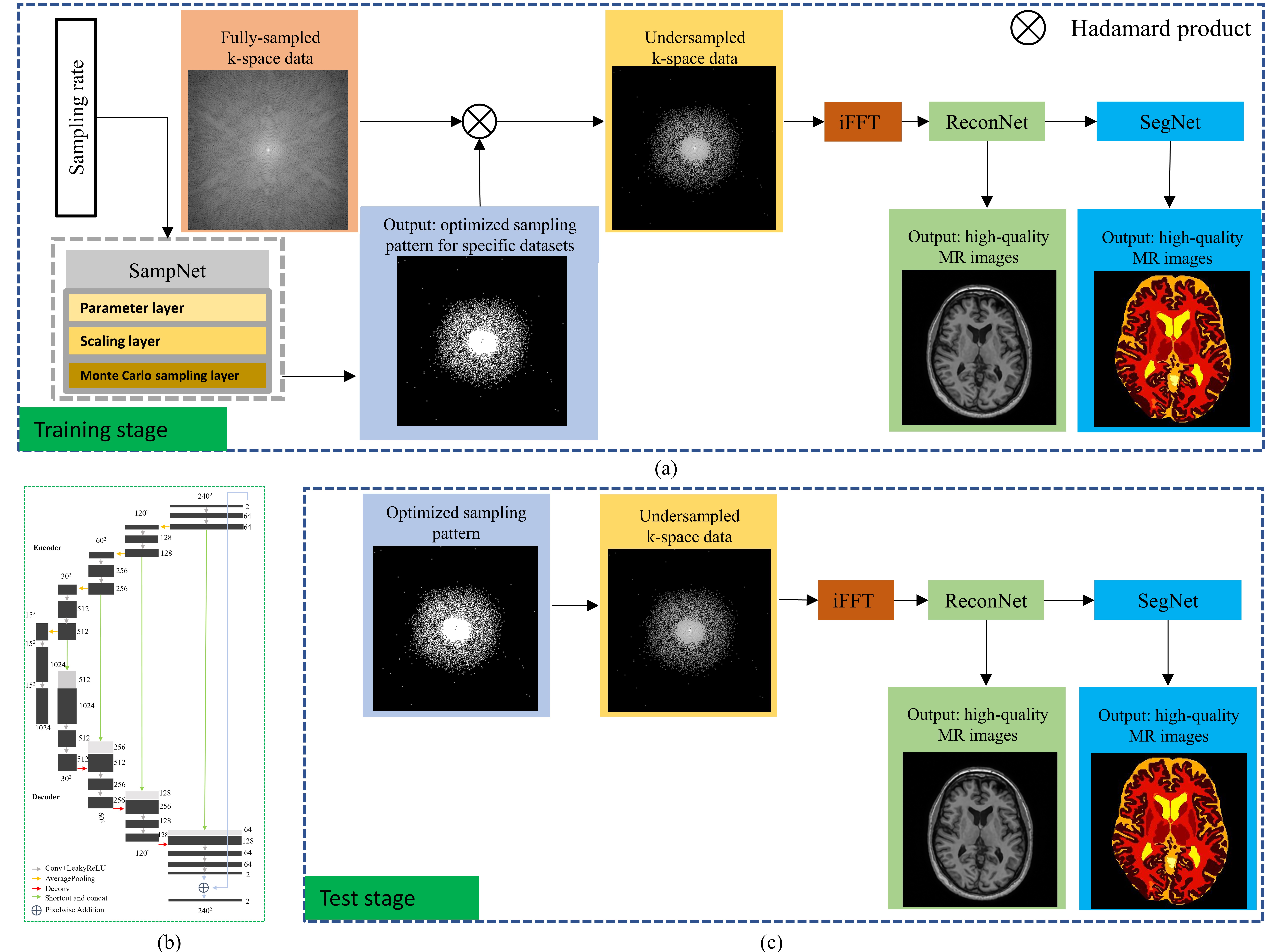}
\caption{Overview of the proposed joint learning network framework: SemuNet.(a) Training stage; (b) An encoder-decoder architecture;(c) Test stage.} \label{fig1}
\end{center}
\end{figure}
The overview of the proposed joint learning network framework is illustrated in Fig.~\ref{fig1}a and Fig.~\ref{fig1}c. It can be seen that the networks in training and testing stages are different. During the training stage, since we need to learn the sampling pattern for the specific dataset with fully-sampled $k$-space data as labels, the whole framework has three parts. During the testing stage, since we can directly use the optimized sampling pattern to acquire the undersampled $k$-space data, SampNet is abandoned and the undersampled $k$-space data is fed into the ReconNet and SegNet in sequence. Finally, the estimated reconstruction and segmentation probability map are obtained.

\subsubsection{Training Strategy.} At the beginning of training stage, the whole network is initialized randomly. The cascaded modules are trained in an end-to-end manner, which updates the weights of three modules simultaneously using backpropagation. The reason to adopt such training strategy is to guarantee the learned sampling pattern can acquire the useful information as more as possible for the subsequent reconstruction and segmentation tasks. More specifically, the proposed SemuNet can be easily adapted to different clinical tasks and we can substitute the SegNet with any other task networks. Our approach not only facilitates the training effort while imposing ReconNet to fit clinical tasks and keeping SegNet performing accurately for undersampled MR images but also enables SampNet to learn more clinically useful features from the $k$-space data.
\subsubsection{Loss function.} For MR images reconstruction, $L_1$ norm is adopted as the loss function:
\begin{equation}\label{eq8}
\mathcal{L}^{Recon} = || R_{\theta}(\textbf{F}^{-1}S_{\textbf{T}_c}(\textbf{x})\odot\textbf{F}\textbf{x})- \textbf{x}||_1 
\end{equation}

Cross-entropy loss is utilized for the SegNet:
\begin{equation}\label{eq9}
\mathcal{L}^{Seg} = -\frac{1}{NC}\sum^N_{n=1}\sum^C_{c=1}t^{gt}_{c,n}\ln{p_{c,n}}
\end{equation}
\begin{figure}
\begin{center}
\includegraphics[width=0.8\textwidth]{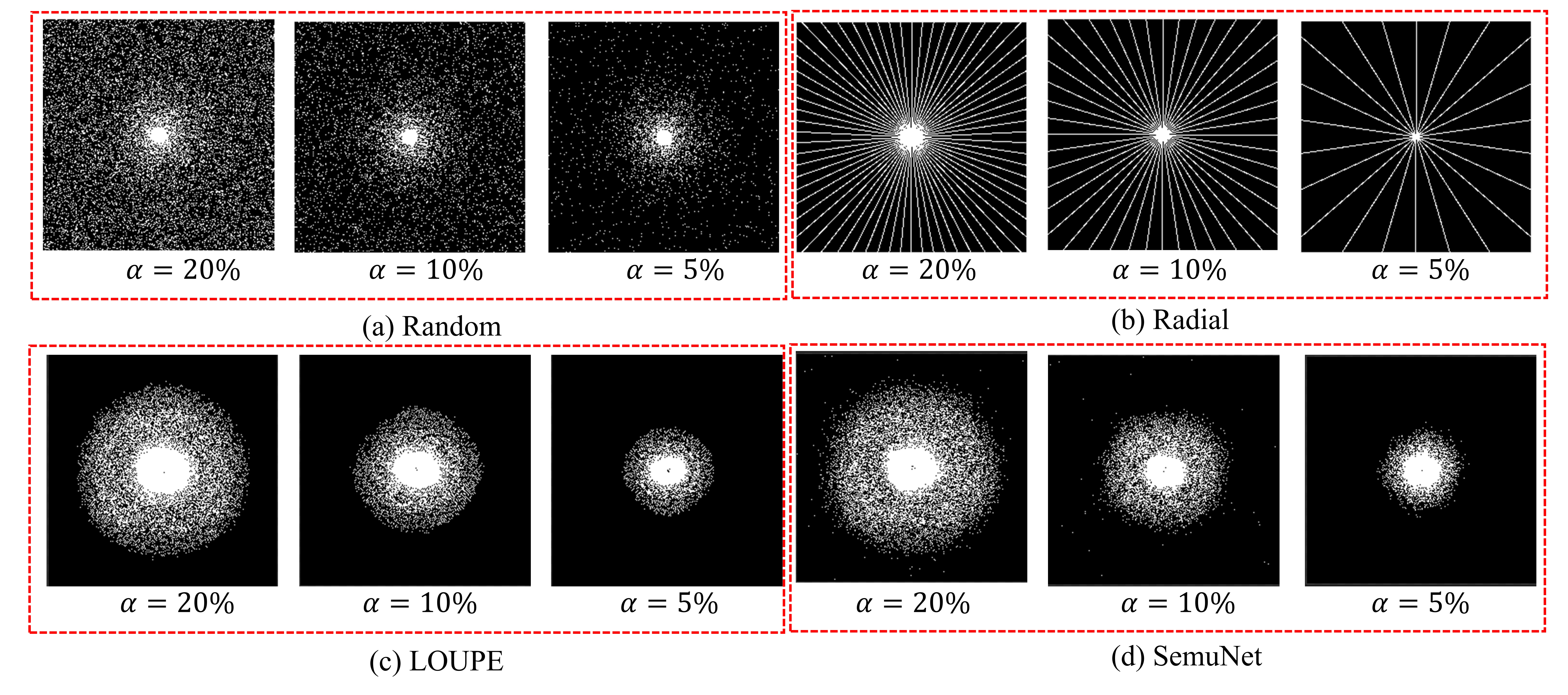}
\caption{Sampling patterns at different undersampling rates for MRB test dataset. (a) random and (b) radial patterns for Baseline, Liu et al.~\cite{ref_article2} and MD-Recon-Net~\cite{ref_article16}; (c) pattern learned by LOUPE~\cite{ref_article8};  (d) pattern learned by our SemuNet. }
 \label{fig2}
\end{center}
\end{figure}\noindent
for $C$ brain tissues class labels and $N$ pixel number of an image, 
where $t_{c,n}^{gt}$ is the pixel-level target label and $p_{c,n}$ is the pixel-level Softmax segmentation probability for the $c^{th}$ class of the $n^{th}$ pixel. Then the hybrid loss function for the proposed joint learning network is formulated as:
\begin{equation}\label{eq10}
\mathcal{L} = \mathcal{L}^{Recon} + \lambda \mathcal{L}^{Seg}
\end{equation}

\section{Experiments and Discussion}
\subsection{Experimental Details}
\subsubsection{Dataset and baselines.} The brain dataset from the Grand Challenge on MR Brain Image Segmentation workshop (MRB)~\cite{ref_article17} is used to evaluate the proposed method. The dataset is acquired using 3.0T MRI scan and consists of five patients. The dataset of each patient is provided with four MRI modalities: T1, T1-1mm, T1-IR and T2-FLAIR with size of $240\times240\times48$. The brain tissues of each patient are manually labeled with seven types of tissue (T1): cortical gray matter, basal ganglia, white matter, cerebrospinal fluid in the extracerebral space, ventricles, cerebellum, and brainstem. In our experiment, four T1 datasets are used for training and the remaining one for testing. 
\subsubsection{Experiment setup.} All implementations are based on Pytorch. All models are trained using one Quadro RTX 8000 GPU and the batch size is set to 12. The hyperparameter configuration of both ReconNet and SegNet are given in Fig.~\ref{fig1}b. Uniform random initialization is used for SampNet and Xavier initialization for ReconNet and SegNet. The whole SemuNet is trained for 600 epochs. After that, the ReconNet and SegNet are fine-tuned for additional 500 epochs. ADAM~\cite{ref_article18} optimizer is adopted with an initial learning rate of $10^{-4}$. $\lambda$ is empirically set to $10^{-1}$. 
\subsubsection{Baseline.} Two basic variants of our SemuNet framework are built: (1) Baseline = fixed pattern + ReconNet + SegNet; and (2) LOUPESeg = LOUPE + SegNet. LOUPE is a recently proposed sampling pattern learning model driven
\begin{figure}
\begin{center}
\includegraphics[width=0.8\textwidth]{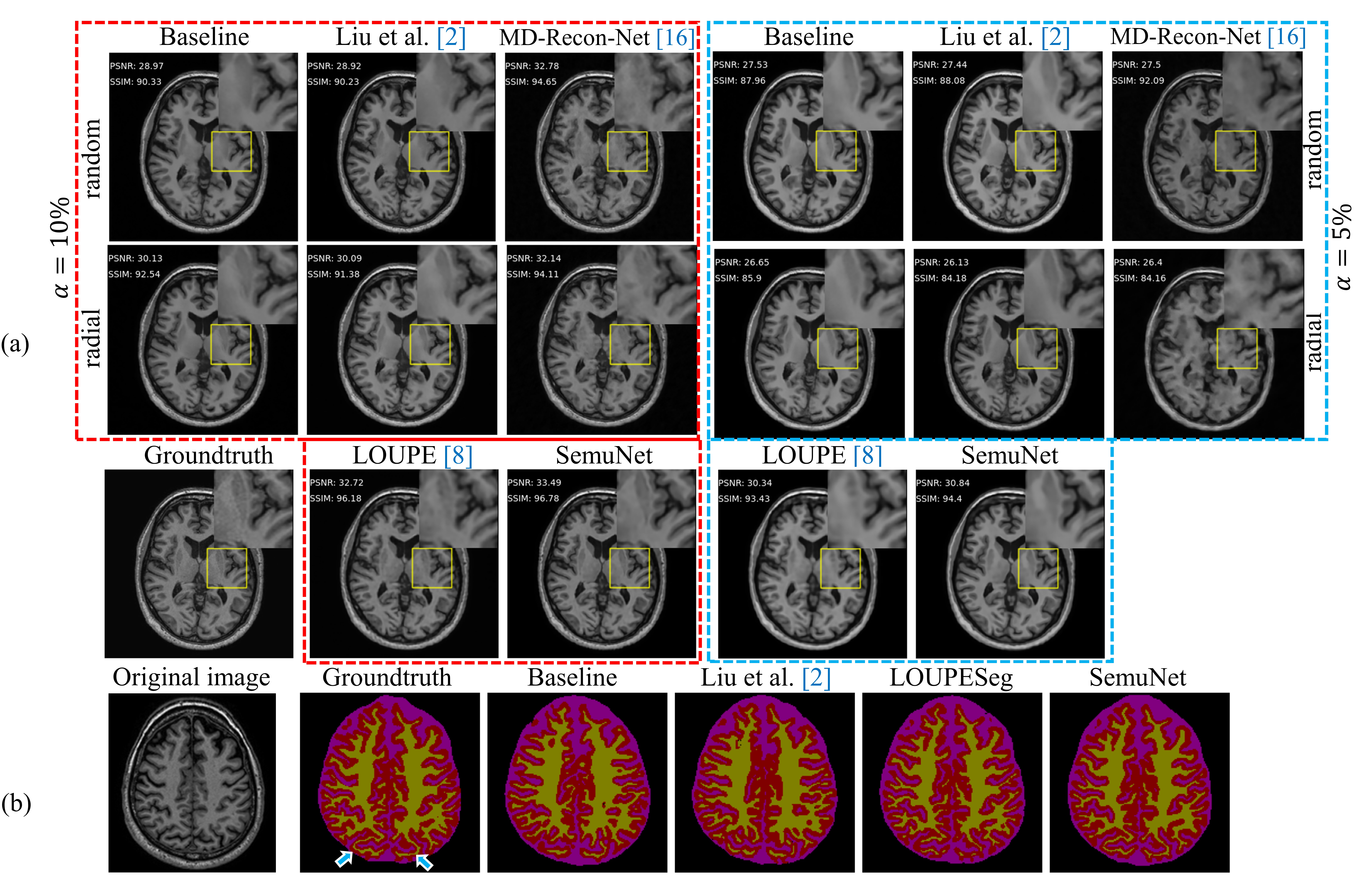}
\caption{(a) Reconstruction results of competing methods at different acceleration factors for MRB test dataset; (b) A brain tissue segmentation example from MRB test dataset ($\alpha=5\%$) using different methods, the $k$-space data for Baseline and Liu et al.~\cite{ref_article2} are undersampled by random pattern.} \label{fig3}
\end{center}
\end{figure}
by reconstruction~\cite{ref_article8}. We first trained LOUPE  with high quality MR images and then SegNet is trained with the data generated by LOUPE. PSNR and SSIM are adopted as quantitative metrics.

\subsection{Experiments Results}
To validate the performance of the proposed SemuNet, we separately evaluate the results of reconstruction and segmentation.

For reconstruction, we compare the proposed SemuNet with the following methods: (1) Baseline (only use its reconstruction result); (2) Liu et al.~\cite{ref_article2} with a fixed pattern (only use its reconstruction result); (3) LOUPE;  (4) MD-Recon-Net~\cite{ref_article16} (a recently proposed dual-domain reconstruction network) with a fixed pattern. The learned trajectories for LOUPE and SemuNet, and the fixed patterns used in Baseline, Liu et al.~\cite{ref_article2} and MD-Recon-Net~\cite{ref_article16} are shown in Fig.~\ref{fig2}a, respectively.

\begin{table}
\centering
    \caption{MRI reconstruction results using PSNR (dB) and SSIM ($\%$) of different methods on the MRB test dataset. The best result is shown in bold.}\label{tab1}
\scalebox{0.8}{
\begin{tabular}{cccccccccc}
\toprule
\multirow{2}{*}{$\alpha$} & \multirow{2}{*}{Metric} & \multicolumn{2}{c}{Baseline} & \multicolumn{2}{c}{Liu et al.~\cite{ref_article2}} & \multicolumn{2}{c}{MD-Recon-Net~\cite{ref_article16}} & LOUPE~\cite{ref_article8} & SemuNet \\ 
\cmidrule(r){3-4} \cmidrule(r){5-6}  \cmidrule(r){7-8}  \cmidrule(r){9-9} \cmidrule(r){10-10} 
 &  & Radial & Random & Radial & Random & Radial & Random & Learned & Learned \\ \midrule
\multirow{2}{*}{20\%} & PSNR & 36.30 & 32.69 & 36.17 & 32.81 & \textbf{39.44} & 39.34 & 38.82 & 39.24 \\ \cmidrule{2-10}
 & SSIM & 96.67 & 94.01 & 96.24 & 93.76 & 98.09 & 97.80 & 98.09 & \textbf{98.56} \\ \midrule
\multirow{2}{*}{10\%} & PSNR & 31.30 & 30.26 & 31.14 & 30.31 & 32.62 & 33.95 & 33.63 & \textbf{34.30} \\ \cmidrule{2-10}
 & SSIM & 90.27 & 90.69 & 90.93 & 90.73 & 93.59 & 94.47 & 95.19 & \textbf{96.47} \\ \midrule
\multirow{2}{*}{5\%} & PSNR & 27.45 & 28.95 & 27.26 & 28.87 & 26.83 & 30.12 & 30.96 & \textbf{31.20} \\ \cmidrule{2-10}
 & SSIM & 85.00 & 88.46 & 84.45 & 88.54 & 85.13 & 92.00 & 91.31 & \textbf{93.16}\\ \bottomrule
\end{tabular}}
\end{table}

\begin{table}
\centering
\caption{DSC ($\%$) of different methods on the MRB test dataset The best result is shown in bold.}\label{tab2}
\scalebox{0.8}{
\begin{tabular}{ccccccc}
\toprule
\multirow{2}{*}{$\alpha$} & \multicolumn{2}{c}{Baseline} & \multicolumn{2}{c}{Liu et al.~\cite{ref_article2}} & \multirow{2}{*}{LOUPE-Seg} & \multirow{2}{*}{SemuNet} \\
\cmidrule(r){2-3} \cmidrule(r){4-5}
& Random & Radial & Random & Radial &  &  \\ \midrule
20\% & 70.65 & 73.91 & 71.64 & 73.77 & 76.19 & \textbf{76.79} \\ \midrule
10\% & 68.3 & 70.48 & 67.73 & 70.92 & 72.91 & \textbf{75.08} \\ \midrule
5\% & 66.6 & 64.54 & 64.66 & 63.65 & 70.97 & \textbf{72.45}\\ \bottomrule
\end{tabular}}
\end{table}
In Fig.~\ref{fig3}a, one typical slice reconstructed using different methods is chosen for visual comparison. It can be observed that the proposed SemuNet achieves the minimal reconstruction error and preserves more details than other methods which can be confirmed in the magnified regions. The average values of the quantitative metrics on the 48 test data (from one patient) are listed in Table~\ref{tab1}. It is noticed that our method achieves the highest scores in most situations, which can be seen as a powerful evidence of that integrating sampling learning and segmentation tasks can efficiently improve the reconstruction performance.

As for segmentation, we compare our method with several methods: (1) Baseline; (2) LOUPESeg; and (3) Liu et al.~\cite{ref_article2}. The results of one representative slice are demonstrated in Fig.~\ref{fig3}b. Each tissue is marked with a different color. It can be observed that the proposed SemuNet provides the most approximate visual result to the ground truth. Dice Similarity Coefficient (DSC)~\cite{ref_article19} is adopted as the quantitative metric and the results are list in Table~\ref{tab2}. The quantitative results are consistent with the subjective evaluation, which confirm that introducing both sampling and reconstruction learning into segmentation network can further increase the accuracy. It is worth noting that the Baseline and Liu et al.~\cite{ref_article2} obtain much lower accuracy than other methods as shown in both Fig.~\ref{fig3}b and Table~\ref{tab2}, which shows the merit of undersampled MR image reconstruction with sampling learning as a preprocessing step for segmentation task. When we only apply sampling pattern learning without considering segmentation task, it also fails to achieve the highest accuracy since the reconstruction does not fully explore the latent features transferred from the segmentation task.
\section{Conclusion}
Sampling pattern learning is an important problem for MR imaging. With the recent developments of fast MRI in the industry, sampling pattern learning technique that takes both reconstruction and analysis tasks into account are of great significance. In this paper, a joint learning framework SemuNet, is proposed to integrate sampling pattern learning, reconstruction and segmentation into a unified network. The results demonstrate the joint learning strategy can benefit all the tasks from each other. In the future work, more datasets will used for evaluation and different analysis tasks will be considered.

\clearpage
\appendix
\section{Supplementary Material}
\setcounter{figure}{0} 
\begin{figure}
\begin{center}
\includegraphics[width=1\textwidth]{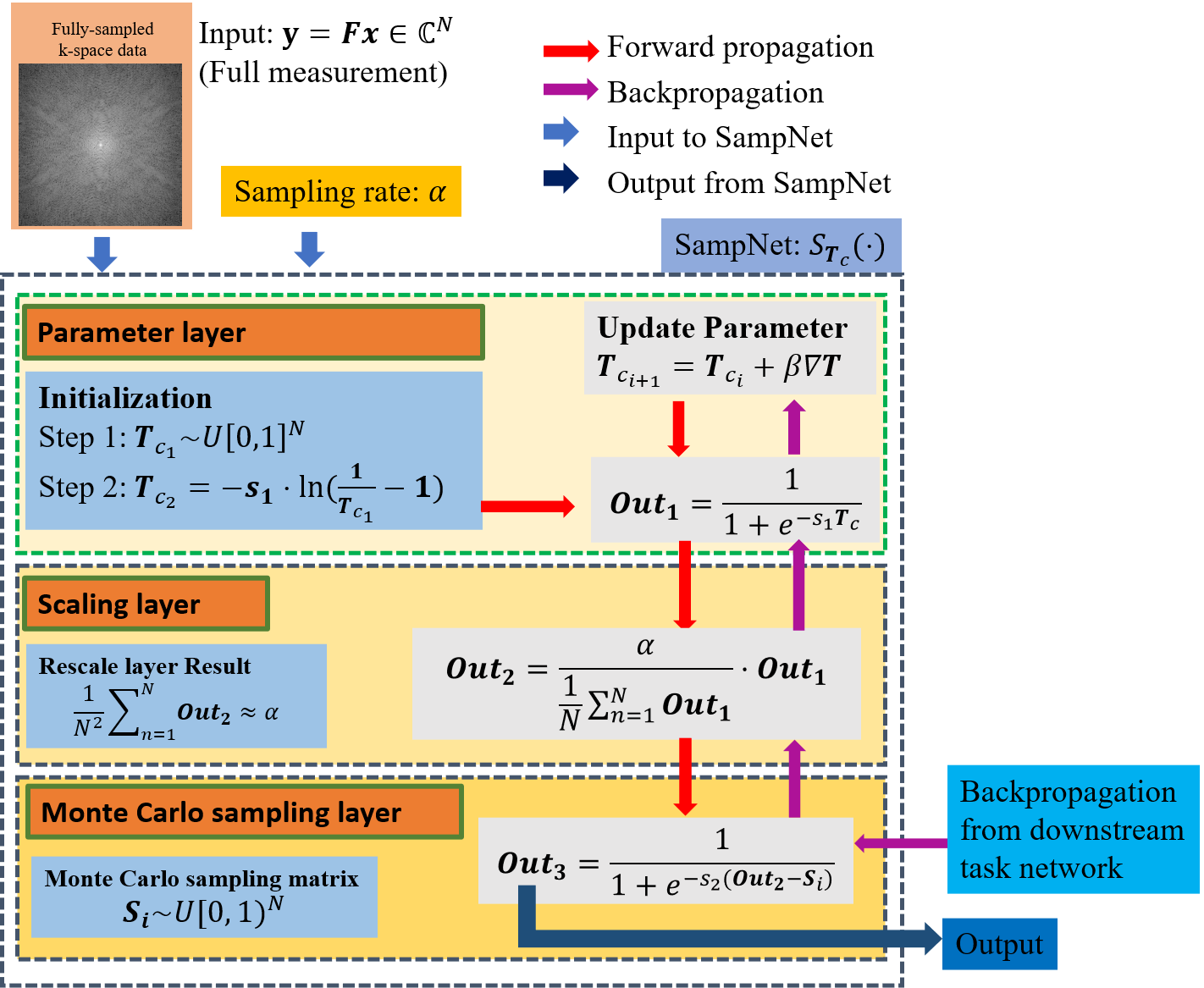}
\caption{The detailed architecture of SampNet. The proposed SampNet is composed of three layers, i.e., parameter layer, scaling layer and Monte Carlo sampling layer. At first, $\textbf{T}_c$ is initialized using uniform distribution on the interval $[0,1)$. Then $\textbf{T}_c$ is mapped to real number set with inverse Sigmoid function. In the parameter layer, it performs no-linear transform using parametrized Sigmoid function. Next, we rescale the mean of the input matrix to a predetermined sampling rate $\alpha$ in scaling layer. In the last layer, Monte Carlo sampling is approximately realized and the parametrized Sigmoid function is also applied on the result.}

\end{center}
\end{figure}\noindent
\end{document}